\documentclass[prb,aps,twocolumn,showpacs,superscriptaddress]{revtex4-1}
\usepackage{amsfonts,amsmath,bm,multirow}
\usepackage[OT4]{fontenc}

\usepackage{graphicx}
\usepackage{xcolor}
\usepackage{epstopdf}
\usepackage{dsfont}
\usepackage{tabularx}
\usepackage{hf-tikz}
\usepackage{physics}
\usepackage{hyperref}
\usepackage{placeins}

\newcommand{\kp}{\bm{k}\!\vdot\!\bm{p}}
\newcommand{\rr}{\bm{r}}
\newcommand{\kk}{\bm{k}}
\newcommand{\eps}{\epsilon}
\newcommand{\exx}{\epsilon_{xx}}
\newcommand{\eyy}{\epsilon_{yy}}
\newcommand{\ezz}{\epsilon_{zz}}
\newcommand{\exy}{\epsilon_{xy}}
\newcommand{\eyz}{\epsilon_{yz}}
\newcommand{\ezx}{\epsilon_{zx}}

\definecolor{bred}{HTML}{e31a1c}
\definecolor{bgreen}{HTML}{33a02c}
\definecolor{bblue}{HTML}{1f78b4}

\definecolor{armygreen}{rgb}{0.29, 0.33, 0.13}

\newcommand{\refa}{$^{\color{blue} a}$}
\newcommand{\refb}{$^{\color{bgreen} b}$}

\newcommand{\fitted}{$^{\color{red} f}$}

\renewcommand{\cp}{{\mathrm{c.p.}}}

\begin{document}
	
	\title{The importance of second order deformation potentials in modeling of InAs/GaAs nanostructures}
	\author{Krzysztof Gawarecki}
	\email{Krzysztof.Gawarecki@pwr.edu.pl}
	\affiliation{Department of Theoretical Physics, Faculty of Fundamental Problems of Technology, Wroc\l aw University of Technology, Wybrze\.ze Wyspia\'nskiego 27, 50-370 Wroc\l aw, Poland}
	\author{Micha{\l} Zieli{\'n}ski}
	\affiliation{Instytut Fizyki UMK, Grudzi\k{a}dzka 5, 87-100 Toru\'n, Poland}
	
	\begin{abstract}
		Accurate modeling of electronic properties of nanostructures is a challenging theoretical problem. Methods making use of continuous media approximation, such as $\kp$, sometimes struggle to reproduce results obtained with more accurate atomistic approaches. 
		On the contrary, atomistic schemes generally come with a substantially larger cost of computation.
		Here, we bridge between these two approaches by taking 8-band $\kp$ method augmented with non-linear strain terms fit to reproduce $sp^3d^5s^*$ tight-binding results. 
		We illustrate this method on the example of electron and hole states confined in quantum wells and quantum dots of photonics applications relevant InAs/GaAs material system, and demonstrate a good agreement of a non-linear $\kp$ scheme with empirical tight-binding method. We discuss limits of our procedure as well as provide non-linear 8-band $\kp$ parameter sets for InAs and GaAs. Finally, we propose a parameterization for effective term used to improve the accuracy of the standard effective mass method.
	\end{abstract}
	
	\maketitle
	
	\section{Introduction}
	\label{sec:intr}
	The realm of nanostructures is a very broad family of systems varying from colloidal nanocrystals,~\cite{delerue-book} self-assembled~\cite{bimberg-book,michler,michler2017quantum,tartakovskii_2012} and nanowire quantum dots~\cite{bjork,borgstrom-nwqd} and quantum dashes~\cite{dery2004nature,Khan2014}, nanowires,~\cite{vls-wagner,dalacu-selective} quantum wells, single dopants in silicon,~\cite{Zwanenburg_RMP13,Usman2016} and more. 
	One of the key points necessary for understanding of physical properties of nanostructures involves accurate and computationally efficient theoretical studies of their electronic structure.
	Since calculations for nanostructures may involve multi-million computational boxes, nanostructures are usually beyond the reach of current ab-initio methods.~\cite{szabo-ostlund,siesta} Therefore, various semi-empirical approaches are typically employed. Moreover, usually researchers opt for one of two apparently opposite schemes. One group of methods is based on the assumption of underlying continuous media. This set involves in particular one-band effective mass approximation, as well as different flavours of multi-band (e.g. 4-, 8- or 14-) $\kp$ method.~\cite{stier-grundmann-bimberg,kp-schliwa,Schliwa111,Schulz111,kp14} These approaches combine very high-computational efficiency with unambiguous parameter sets, that can be obtained directly from bulk properties.
	Moreover, the $\kp$ approaches turned out to be highly successful in understanding of main spectral~\cite{pryor-zunger,wojs,weidong-localization,sheng-cheng-prb2005,ardelt16,climente08,leger,marynski2013electronic,gawelczyk2017exciton} and spin-related\cite{andlauer09,Gawarecki2018a,Gawelczyk2018a} features for a broad group of nanostructures. 
	
	There is however a second group of widely acknowledged, so-called ''atomistic'' methods based on explicit accounting for low, atomistic symmetry of nanostructures, which are defined in calculations atom-by-atom. 
	Notably, these~\footnote{We note that there are some intermediate schemes in particular the EBOM method, which however does not have atomistic spatial resolution.} include empirical pseudopotential method,~\cite{Zunger-EPM2,bester-zunger-nair,bester-zunger-p-shell,singh-bester-ordering,singh-bester-lower} and empirical tight-binding method.~\cite{slater-koster,vogl,jancu,santoprete03,saito-arakawa,schultz12,schulz-schum-czycholl,alloyed-qd,boykin-fit,voisin-fit,jaskolski06,jaskolski04,zielinski12,Zielinski2013}
	Atomistic approaches often provide more accurate results (such as e.g. the magnitude of the bright exciton splitting~\cite{bayer-eh,takagahara,Swiderski2017,Zielinski2013}) and are methods of choice when one deals with effect of alloying,~\cite{rand-mlinar-zunger,singh-bester-ordering,singh-bester-lower,Zielinski-natural,Swiderski2017,zielinski12,dash-mrowinski} low-shape symmetry and faceting,~\cite{bryant-prl,bryant-prb,Zielinski.PRB.2015,wojnar2014strain} monolayer-thin sizes~\cite{Diaz.PhysRevB.75.245433} or sections of semiconductor,~\cite{zhang,nika-cpqd,our-cpqd} and atomistically sharp interfaces.~\cite{Zwanenburg_RMP13,maaike-strain,nw-sharp} 
	Atomistic methods are however usually much more computationally demanding,~\cite{alloyed-qd,zielinski-multiscale,rozanski-zielinski,rozanski-zielinski-cpc} with often complicated and non-trivial schemes for semi-empirical parameters fitting.~\cite{alloyed-qd,boykin-fit,voisin-fit,tan-fit}
	
	The properties of semiconductor nanostructures are strongly modified by the presence of strain, which is inevitable for a system composed of lattice mismatched materials.
	In atomistic methods, strain is accounted for by displacements of the atomic positions which alters the bond lengths and bond angles. On the other hand, in continuous media approaches, strain is represented by macroscopic strain tensor field. In a standard way, strain enters the $\kp$ model via well-established Bir-Pikus Hamiltonian, containing strain-tensor elements in linear order\cite{Bir1974, Trebin1979, bahder90}. This approach accurately describes the band structure, if strain is low or moderate. To describe nonlinear effects (visible in DFT simulations for InAs and GaAs materials\cite{Kadantsev2012}) which are relevant at stronger strain, higher order strain terms would be needed.  Despite the second order scheme was proposed\cite{Suzuki1974} no reliable parameterization is available.
	
	In this work we aim to bridge between these two seemingly excluding ways of calculations and improve $\kp$ results considerably by accounting for second-order strain effects.
	We thus emphasize and study the importance of nonlinear strain effects on the spectral properties of nanostructures. 
	To this end, we go beyond the standard Bir-Pikus Hamiltonian and implement quadratic strain terms with the second order deformation potentials\cite{Suzuki1974}. By comparison to the $sp^3d^5s^*$ tight binding method, we find the relevant parametrization for InAs and GaAs bulk materials. Then, using $8$-band $\kp$ Hamiltonian (supplemented by the 2nd order strain terms), we calculate the electron and hole energy levels in classes of quantum well (QW) and quantum dot (QD) structures. We show, that single particle energies are strongly affected by nonlinear terms related to the biaxial strain. We also demonstrate that, in the case of electron in the QW and QD, an excellent agreement between the $8$-band $\kp$ and the $sp^3d^5s^*$ tight-binding model can be achieved if second order strain terms are taken into account. These findings are not only important from the fundamental point of view, but also to properly predict on the the optoelectronic or nanophotonic device characteristics, especially when calculating the excitonic properties in single quantum dots of this material system broadly considered as efficient sources of single or entangled photon states also at the telecommunication spectral range of 1.3 - 1.55 $\mu$m.
	
	The paper is organized as follows. In Sec.~\ref{sec:bulk}, we describe the models, calculate the band structure, and fit the material parameters used in further simulations. In Sec.~\ref{sec:nano}, we calculate the electron and hole states in the QW and the QD structures. Finally, Sec.~\ref{sec:concl} contains concluding remarks. Furthermore, the derivation of strain Hamiltonian is presented in the Appendix.
	
	\section{Bulk material}
	\label{sec:bulk}
	
	The reference band structures for InAs and GaAs are obtained within $sp^3d^5s^*$ tight-binding model\cite{Slater1954}, where we took the material parameters from Ref.\cite{Jancu1998}. To account for the strain, we manipulate the atomic positions in the primitive cell. The bond length variations enter the Hamiltonian via the orbital-dependent exponents  rescaling the two-center integrals (a generalization of the Harrison law).\cite{Jancu1998,harrison-big-book} 
	Inter-atomic (hopping) matrix elements $t$ are thus modified as:
	\begin{align*}
	t=t_0\left(\frac{d_0}{d}\right)^\eta,
	\end{align*}
	where $t_0$ ($t$) is unstrained (strained) hopping matrix element, $d_0$ ($d$) is unperturbed (perturbed) bond-length,$\eta$ is scaling matrix element whose magnitude is fitted to reproduce bulk deformation potentials.~\cite{Jancu1998,boykin-fit}
	While the hydrostatic strain alters only the bond lengths; the uniaxial, biaxial and shear strain change the bond angles as well. 
	Furthermore, the uniaxial and biaxial strain lead to the energy splitting in the atomic $d$ shell. The  $d_{xy}$, $d_{yz}$ and $d_{zx}$ orbital on-site matrix elements are then given by\cite{Jancu1998}
	$$
	E_{xy} = E_{d} + 2 b_{d} \qty[ \ezz - (\exx + \eyy)/2 ],
	$$
	where $E_{yz}$, $E_{zx}$ are given by cyclic permutations of indices; $E_{d}$ represents the unperturbed $d$-shell energy, $b_{d}$ is a material-dependent parameter, and  $\eps_{ij}$ are the strain tensor components. 
	We implemented the matrix elements using the form described in Ref.\cite{zielinski12}, where the strain tensor components are expressed in terms of the bond lengths and direction cosines.
	We note here that strain related power-dependence of hopping matrix elements, as accounted by Harrison law, is inherently non-linear and may lead to non-linear bulk bands evolution under external strain, in particular for larger strain as present in InGaAs nanostructures.
	
	The second approach, which we utilize in the paper is the $8$-band $\kp$ model, where the Hamiltonian explicitly contains $\Gamma_\mathrm{6c}, \Gamma_\mathrm{8v}$ and $\Gamma_\mathrm{7v}$ blocks corresponding to the irreducible representations of the $T_d$ symmetry point group\cite{mayer91,Winkler2003}.
	According to the invariant expansion scheme\cite{Luttinger1955},  the Hamiltonian can be expressed in terms of the matrices representing crystal symmetry invariants. Then, the kinetic part of the Hamiltonian is written\cite{Winkler2003}
	\begingroup
	\allowdisplaybreaks
	\begin{subequations}
		\begin{align*}
		H_{k,\mathrm{6c6c}} &= E_{g} + A'_{\mathrm{c}} k^{2},\\
		H_{k,\mathrm{8v8v}} &= -\frac{\hbar^{2}}{2m_{0}} \left \{ \gamma'_{1}  k^{2} - 2 \gamma'_{2} \left [  \left (  J^{2}_{x}  - \frac{1}{3} J^{2}  \right ) k^{2}_{x}  + \cp  \right ] \right . \nonumber \\
		&\phantom{=} - 4 \gamma'_{3} \left [ \{J_{x},J_{y}\}  \{k_{x},k_{y}\}  + \cp  \right ]  \bigg \} \nonumber\\
		&\phantom{=} -i \frac{\hbar^{2}}{m_{0}} \qty\Big(  \kappa' [k_{x} , k_{y} ]  J_{z} + q' [k_{x} , k_{y} ] J^{3}_{z}   + \cp ) \nonumber\\
		&\phantom{=} + \frac{2}{\sqrt{3}} C_{k} \left[ \{ J_{x},J^{2}_{y} - J^{2}_{z}  \} k_x + \cp  \right], \\
		H_{k,\mathrm{7v7v}} &= -\Delta_{0} - \frac{\hbar^{2}  k^{2}}{2m_{0}} \gamma'_{1} \nonumber \\
		&\phantom{=}  -i \frac{\hbar^{2}}{m_{0}} \qty\Big( \kappa' [k_{x},  k_{y}]  \sigma_{z} + \cp ),\\
		H_{k,\mathrm{6c8v}} &= \sqrt{3} P \bm{T} \cdot \kk + i\sqrt{3} B_{\mathrm{8v}}^{+} \qty\Big( T_{x}  \{ k_{y} , k_{z} \} +\cp ) \nonumber \\
		& \quad +\frac{\sqrt{3}}{2}  B_{\mathrm{8v}}^{-} (T_{xx}-T_{yy})  \left ( \frac{2}{3} k^2_{z} -  \frac{1}{3}k^2_{x} - \frac{1}{3}k^2_{y} \right )\nonumber \\
		& \quad- \frac{\sqrt{3}}{2}  B_{\mathrm{8v}}^{-} T_{zz} (k^2_{x} - k^2_{y}),\\
		H_{k,\mathrm{6c7v}} &= - \frac{1}{\sqrt{3}} \qty\Big[ P \bm{\sigma} \cdot \kk + i B_{\mathrm{7v}} ( \sigma_x \{k_y,k_z\} + \cp)   ],\\
		H_{k,\mathrm{8v7v}} &= -\frac{\hbar^{2}}{2m_{0}} \left[ -6 \gamma'_{2} (U_{xx} k^{2}_{x} + \cp ) \right . \nonumber \\
		& \phantom{=}   -12 \gamma'_{3} (U_{xy} \{ k_{x} , k_{y} \} + \cp ) \big ] \nonumber \\
		& \phantom{=} - i \frac{3 \hbar^{2}}{2 m_{0}} \qty\Big( \kappa' U_{z} [k_{x}, k_{y}] + \cp ) \nonumber \\
		& \phantom{=} - i \sqrt{3} C_{k} \left ( U_{yz} k_{x} + \cp \right ),
		\end{align*}
	\end{subequations}
	\endgroup
	where $E_{g}$ is the energy gap, $\Delta_{0}$ describes the spin-orbit coupling, $P$ is a parameter proportional to the interband momentum matrix element, $m_{0}$ is the free electron mass, $A'_{\mathrm{c}}$ accounts for the remote bands contributions to the electron effective mass, $\gamma'_{i}$ are modified Luttiner parameters, $\kappa' = -\frac{1}{3} (\gamma'_{1} - 2 \gamma'_{2} - 3 \gamma'_{3} +2) $, $q'$ is an anisotropy parameter, $B_{\mathrm{8v}}^{\pm}, B_{\mathrm{7v}}$ are parameters related to the Dresselhaus spin-orbit coupling, $\{A,B\} = \frac{1}{2} (AB + BA)$, ,,$\cp$" denotes cyclic permutations, $\sigma_{i}$ are the Pauli matrices, matrices $J_i$ are related to the $j=3/2$ representation of angular momentum, $T_i$ are matrices connecting the $j=1/2$ representation to the $j=3/2$, $T_{ij} = T_{i} J_j + T_{j} J_i$, $U_{i} = T^{\dagger}_{i}$, and $U_{ij} = T^{\dagger}_{ij}$. The explicit definitions of the matrices are given in Refs.\onlinecite{mayer91,Winkler2003,Trebin1979}. 
	
	\begin{figure*}[!tb]
		\begin{center}
			\includegraphics[width=7in]{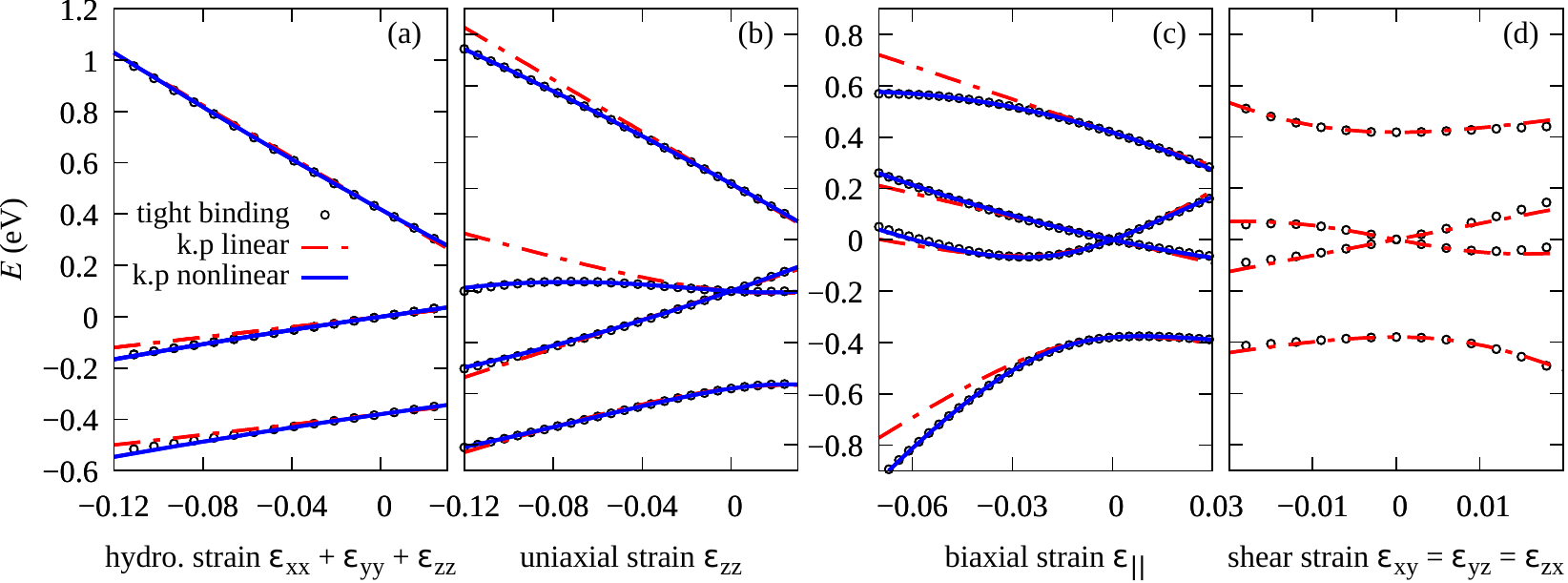}
		\end{center}
		\caption{\label{fig:edges}\textcolor{gray}({Color online) Band edges of InAs as a function of strain. In the case of linear strain approximation, we take the deformation potentials from Ref.\cite{vurgaftman01}} (see Table\ref{tab:dps}). }
	\end{figure*}
	Within the standard Bir-Pikus model, strain tensor elements enters the Hamiltonian in linear order\cite{Winkler2003,bahder90}. At $\kk = 0$, this is given by
	\begingroup
	\allowdisplaybreaks
	\begin{subequations}
		\begin{align*}
		H^{\mathrm{(1)}}_{\mathrm{str},\mathrm{6c6c}} &=  a^{\mathrm{(1)}}_{\mathrm{c}} \Tr{\epsilon},\\
		H^{\mathrm{(1)}}_{\mathrm{str},\mathrm{6c8v}} &=  i \sqrt{3} C_{2} (T_{x} \eyz + \cp),\\
		H^{\mathrm{(1)}}_{\mathrm{str},\mathrm{6c7v}} &=  -\frac{i} {\sqrt{3}} C_{2} (\sigma_{x} \eyz + \cp),\\
		H^{\mathrm{(1)}}_{\mathrm{str},\mathrm{8v8v}} &=  a^{\mathrm{(1)}}_{\mathrm{v}} \Tr{\epsilon} - b^{(1)}_{\mathrm{v}} \left[ \left( J^{2}_{x}  - \frac{1}{3} J^{2} \right)\epsilon_{xx} + \cp  \right]  \nonumber\\
		& \phantom{=} - \frac{d^{\mathrm{(1)}}_{\mathrm{v}}}{\sqrt{3}} \left[ 2 \{J_{x},J_{y}\} \epsilon_{xy}  + \cp \right], \label{str1_8v8v}\\
		H^{\mathrm{(1)}}_{\mathrm{str},\mathrm{8v7v}} &=  -3 b^{\mathrm{(1)}}_{\mathrm{v}} \left( U_{xx} \epsilon_{xx} + \cp \right)  \nonumber\\ &\phantom{=} -\sqrt{3} d^{\mathrm{(1)}}_{\mathrm{v}} \left( 2 U_{xy} \epsilon_{xy} + \cp \right),\\
		H^{\mathrm{(1)}}_{\mathrm{str},\mathrm{7v7v}} &=  a^{\mathrm{(1)}}_{\mathrm{v}} \Tr{\epsilon},
		\end{align*}
	\end{subequations}
	\endgroup
	where $a^{\mathrm{(1)}}_{\mathrm{c}}$, $a^{\mathrm{(1)}}_{\mathrm{v}}$, $b^{\mathrm{(1)}}_{\mathrm{v}}$ and $d^{\mathrm{(1)}}_{\mathrm{v}}$ are the deformation potentials\cite{bir61,bahder90}. The term proportional to the $C_2$ parameter results from the inversion asymmetry\cite{Trebin1979}, and provides a channel of spin-orbit coupling.
	
	We go beyond the standard framework and include the terms which are quadratic in strain tensor elements\cite{Suzuki1974} and correspond to the hydrostatic and biaxial strain, while second order terms with shear strain components are neglected. The relevant part of the Hamiltonian in the invariant expansion is then given by
	\begingroup
	\allowdisplaybreaks
	\begin{subequations}	
		\begin{align*}
		H^{\mathrm{(2)}}_{\mathrm{str},\mathrm{6c6c}} &=  a^{\mathrm{\mathrm{(2a)}}}_{\mathrm{c}} (\exx + \eyy + \ezz)^{2} + a^{\mathrm{(2b)}}_{\mathrm{c}} (\exx \eyy + \cp),\\
		H^{\mathrm{(2)}}_{\mathrm{str},\mathrm{8v8v}} &=  a^{\mathrm{\mathrm{(2a)}}}_{\mathrm{v}} (\exx + \eyy + \ezz)^{2} + a^{\mathrm{(2b)}}_{\mathrm{v}} (\exx \eyy + \cp)
		\\ &\phantom{=} - b^{\mathrm{(2a)}}_{\mathrm{v}} \left[ \left( J^{2}_{x}  - \frac{1}{3} J^{2} \right)\epsilon^{2}_{xx} + \cp  \right] \\
		\\ &\phantom{=} - b^{\mathrm{(2b)}}_{\mathrm{v}} \left[ \left( J^{2}_{x}  - \frac{1}{3} J^{2} \right)\eyy\ezz + \cp  \right],\\
		H^{\mathrm{(2)}}_{\mathrm{str},\mathrm{8v7v}} &=  -3 b^{\mathrm{(2a)}}_{\mathrm{v}} \left( U_{xx} \exx^2 + \cp \right) \\
		&\phantom{=} -3 b^{\mathrm{(2b)}}_{\mathrm{v}} \left( U_{xx} \eyy \ezz + \cp \right) ,\\
		H^{\mathrm{(2)}}_{\mathrm{str},\mathrm{7v7v}} &=  a^{\mathrm{\mathrm{(2a)}}}_{\mathrm{v}} (\exx + \eyy + \ezz)^{2} + a^{\mathrm{(2b)}}_{\mathrm{v}} (\exx \eyy + \cp),
		\end{align*}
	\end{subequations}
	\endgroup
	where six additional deformation potentials are introduced. A detailed derivation is presented in the Appendix.

	We calculated band edges of the InAs and GaAs bulk materials using the tight-binding method. Then, the deformation potentials of $8$-band $\kp$ model are fitted to the TB data (see Table \ref{tab:dps} for values). Fig.~\ref{fig:edges}(a) presents the results of band edges in InAs as a function of hydrostatic strain. In this case, the effect of nonlinearity is weak, and the $H^{\mathrm{(1)}}_{\mathrm{str}}$ already provides a good approximation. The results for uniaxial- and biaxial strain are shown in Figs.~\ref{fig:edges}(b,c)].  For the biaxial strain we use the Poisson ratio $\ezz = - (2 C_{12}/C_{11} )\epsilon_{\parallel}$, where $C_{12}$ and $C_{11}$ are the elastic constants. The mismatch between the Bir-Pikus model and tight-binding results becomes significant, however a good agreement can be achieved, if the second order strain terms ($H^{\mathrm{(2)}}_{\mathrm{str}}$) are included. 
		\begin{table}
		\caption{\label{tab:dps}Deformation potentials (in eV), and parameters of the valence force field model, used in the calculations. 
		}
		\begin{tabular}{m{3cm}m{3cm}l}
			& GaAs & InAs \\
			\hline		
			\rule{0pt}{3ex}$a^{\mathrm{(1)}}_{\mathrm{c}}$ & -6.79\fitted (-7.17\refa)& -4.78\fitted (-5.08\refa)\\
			$a^{\mathrm{(1)}}_{\mathrm{v}}$ & 1.84\fitted (1.16\refa) & 1.24\fitted (1.00\refa) \\
			$b^{\mathrm{(1)}}_{\mathrm{v}}$ & -1.85\fitted (-2.0\refa) & -1.77\fitted (-1.8\refa) \\
			$d^{\mathrm{(1)}}_{\mathrm{v}}$ & -4.8\refa & -3.6\refa \\
			\hline
			\rule{0pt}{3ex}$C_{2}$ & 2.5 -- 5.5 (3.3) & 6.0\fitted \\
			\hline
			\rule{0pt}{3ex}$a^{\mathrm{(2a)}}_{\mathrm{c}}$ & -2.71\fitted & -3.40\fitted \\
			$a^{\mathrm{(2b)}}_{\mathrm{c}}$ & 24.6\fitted & 18.1\fitted \\
			$a^{\mathrm{(2a)}}_{\mathrm{v}}$ & -3.56\fitted & -1.56\fitted \\
			$a^{\mathrm{(2b)}}_{\mathrm{v}}$ &  4.81\fitted &  1.07\fitted \\
			$b^{\mathrm{(2a)}}_{\mathrm{v}}$ & -6.38\fitted & -5.95\fitted \\
			$b^{\mathrm{(2b)}}_{\mathrm{v}}$ & -9.23\fitted & -6.64\fitted \\
			\hline 
			\rule{0pt}{3ex}$\widetilde{a}^{(2)}_\mathrm{c}$ & -3.8\fitted & -3.2\fitted \\
			\hline 
			$\alpha$~$(10^3$ dyne $)$ & $41.49$\refb & $35.18$\refb \\
			$\beta$~$(10^3$ dyne $)$ & $8.94$\refb & $5.49$\refb \\
			\toprule
			\multicolumn{3}{p{0.97\linewidth}}{\rule{0pt}{1.em} \fitted Values fitted to the tight-binding band structure.} \\
			\multicolumn{3}{p{0.97\linewidth}}{\rule{0pt}{1.em} \refa Values taken from Ref.\cite{vurgaftman01}. }  \\
			\multicolumn{3}{p{0.97\linewidth}}{\rule{0pt}{1.em} \refb Values taken from Ref.\cite{pryor98b}. } 
		\end{tabular}
	\end{table}
	
	In the case of hydrostatic and uni-/biaxial strain (represented by macroscopic tensor $\hat \eps$), the displacements of atoms in the unit cell are well defined. 
	On the other hand, if the shear strain is taken into account, the situation becomes more complicated. In such case, the atomic positions are also affected by microscopic relative displacements  between the two crystal sublattices. The atomic positions ($\bm{R}_i$) are then given by\cite{Kleinman1962,Yu2010,Niquet2009}
	$$
	\bm{R}_i = (\mathds{1} + \hat\eps )  \bm{R^0}_i \pm \zeta \frac{a}{4} (\eyz,\ezx,\exy),
	$$
	where ($\bm{R^0}_i$) is the initial (in unstrained crystal)  position of the $i$-th atom, $\zeta$ is the Kleinmann parameter, $a$ is the lattice constant, and the sign $\pm$ depends on the sublattice (anion or cation). In general, the value of Kleinman parameter depends on hydrostatic and shear strain in the system\cite{Garg2009}. Due to uncertainty related to $\zeta$ and the number of free parameters (see the Appendix A), we skipped  fitting of the second order shear strain terms. We approximate the value of Kleinman parameter by a constant value with the formula\cite{Steiger2011}
	$$
	\zeta = \frac{\alpha-\beta}{\alpha+\beta},
	$$
	where $\alpha$, $\beta$ are parameters of the Keating VFF model\cite{pryor98b}. 
	The comparison between the methods is given in Fig.\ref{fig:edges}(d). As the fitting of $C_2$ for GaAs requires high shear strain, its value is not estimated satisfactory (see Table~\ref{tab:dps}). Instead, following Ref.\cite{Mielnik-Pyszczorski2018b} we take its value as $3.3$~eV, which was extracted from the experimental data of spin relaxation time\cite{dyakonov86a}.
	Moreover we note that $sp^3d^5s^*$ parametrization by Jancu~\cite{Jancu1998} has limited accuracy for representing shear strains,~\cite{Jancu2007} and a such may not be the best target for $\kp$ fitting.
	
    In the next step, we fit the band structure parameters for unstrained crystal [see Table~\ref{tab:band_param}] for values]. While the TB parametrization~\cite{Jancu1998} for GaAs is in very good agreement with Ref.\cite{vurgaftman01}, the comparison for InAs requires some rescaling of the Luttinger parameters. Moreover, following Ref.\cite{birner11}, to avoid spurious solutions in further calculations in nanostructure, we set $A'_c=1$ and then rescale $E_\mathrm{p}$.
	%
	\begin{table}
		\caption{\label{tab:band_param}Band structure parameters used in the calculations. We show the values of reduced Luttinger parameters (where the contribution from the $\Gamma_{\mathrm{6c}}$ is substracted), which directly enter the Hamiltonian\cite{Winkler2003}.}
		\begin{tabular}{lll}
			\toprule
			& GaAs & InAs \\
			\hline
			\rule{0pt}{3ex}$E_{\mathrm{v}}$ & $0.0$~eV & $0.21$~eV\refa \\
			$E_{\mathrm{g}}$ & 1.519~eV\refa & 0.418~eV\fitted \\
			$E_{\mathrm{p}}$ & $21.5$~eV\fitted  & $19.5$~eV\fitted  \\
			\rule{0pt}{3ex}$P$ & \multicolumn{2}{c}{calculated  from $P = \sqrt{ E_{p}  \hbar^{2} / (2 m_{0} )}$}\\
			$\gamma'_{\mathrm{1}}$ & 2.26\refa & 2.15\fitted  \\
			$\gamma'_{\mathrm{2}}$ & -0.299\refa & -0.325\fitted   \\
			$\gamma'_{\mathrm{3}}$ & 0.571\refa & 0.542\fitted  \\
			$C_{\mathrm{k}}$ & -0.0034~$\mathrm{eV \AA}$\refb & -0.0112~$\mathrm{eV \AA}$\refb \\
			$\Delta_{0}$ & 0.341~eV\refa & 0.38~eV\fitted  \\
			\toprule
			\multicolumn{3}{p{0.97\linewidth}}{\rule{0pt}{1.em} \fitted Values fitted to the tight-binding band structure.} \\
			\multicolumn{3}{p{0.97\linewidth}}{\rule{0pt}{1.em} \refa Values taken from Ref.\cite{vurgaftman01}. } \\
			\multicolumn{3}{p{0.97\linewidth}}{\rule{0pt}{1.em} \refb Values taken from Ref.\cite{Winkler2003}. } \\
		\end{tabular}
	\end{table}
	
	In the case of the (single band) effective mass model, the most important second order strain terms enter the Hamiltonian with a single effective parameter. If we neglect relatively small nonlinearity of the hydrostatic strain, the effective mass Hamiltonian for the electron can be written as
	\begin{align*}
	\widetilde{H}_\mathrm{eff} &= \sum_i k_i \frac{\hbar^2}{2 m'} k_i + E_{\mathrm{c}} + a_\mathrm{c} (\exx + \eyy + \ezz) \\
	&\phantom{=} + \widetilde{a}^{(2)}_\mathrm{c} [(\exx - \eyy)^2 + (\eyy - \ezz)^2 + (\ezz - \exx)^2 ],
	\end{align*}
	where $m'$ is the electron effective mass, $E_{\mathrm{c}}$ is the conduction band (cb.) edge, $ a_\mathrm{c}$ is the standard cb. deformation potential,  $\widetilde{a}^{(2)}_\mathrm{c}$ is the effective parameter related to the biaxial strain in the second order. The values of $\widetilde{a}^{(2)}_\mathrm{c}$ for InAs and GaAs, (where the fitting procedure was optimized for the negative biaxial strain) are given in Table~\ref{tab:dps}.
	
	\section{Nanostructures}
	\label{sec:nano}
	
	In this Section, we utilize the models described previously for the calculations of carrier states in QWs and QDs. We find the strain distribution in the system within standard Keating VFF model\cite{pryor98b}.  The elastic energy is given by
	\begin{align*}
	U &= \frac{3}{16} \sum_{i} \sum_{j}^{\mathrm{NN}(i)}  A_{ij} \left( \bm{r}_{ij}^{2} - d_{ij}^{2} \right)^{2} \\
	& \phantom{=} + \frac{3}{8} \sum_{i} \sum_{j}^{\mathrm{NN}(i)} \sum_{k>j+1}^{\mathrm{NN}(i)}  B_{ijk}  \left( \bm{r}_{ij} \bm{r}_{ik} - d_{ij} d_{ik} \cos{\theta_{0}} \right)^{2},
	\end{align*}
	where $\mathrm{NN}(i)$ denotes nearest neighbors of $i$-th atom, $\bm{r}_{ij}=\bm{r}_{i}-\bm{r}_{j}$ and $d_{ij}$ are actual and idealized (unrelaxed) distances between $i$-th and $j$-th atom,
	$A_{ij}=\alpha_{ij}/d_{ij}^{2}$ with bond-streching constant $\alpha_{ij}$,  $B_{ijk}= (\beta_{ij}  + \beta_{ik} )/ (2 d_{ij} d_{ik})$, here $\beta_{ij}$ is a constant which represents bond bending.
	We use PETSC TAO\cite{petsc} library to minimize the elastic energy of the system via relaxation of the atomic positions. The strain tensor elements are obtained from the direction cosines and the bond lengths, in a way described in Ref.~\cite{zielinski12}. We find the electron and hole states within multimilion atom simulations using the $sp^3d^5s^*$ tight-binding model\cite{Slater1954,zielinski12,Jancu1998}. The passivation of dangling-bonds at the limit of computational domain, has been performed following Ref.\cite{Lee2004}.
	
	For the 8-band $\kp$ simulations, we move from the atomic lattice to the cartesian grid by taking $h_x = h_y = a_{\mathrm{GaAs}}$, $h_z = a_{\mathrm{GaAs}}/2$ mesh size, and averaging the strain and composition over the two cations in each mesh cell. To calculate electron and hole states in a nanostructure within the 8-band $\kp$ model, we perform the standard substitution $k_i = -i \hbar \pdv{x_i}$ in the bulk Hamiltonian. Since all of the material parameters are position dependent, $k_{i}$ does not commute with them and the operator ordering becomes important. The details of the implementation are described in the Appendix of Ref.\cite{Gawarecki2018a}.
	
	\begin{figure}[!tb]
		\begin{center}
			\includegraphics[width=3.5in]{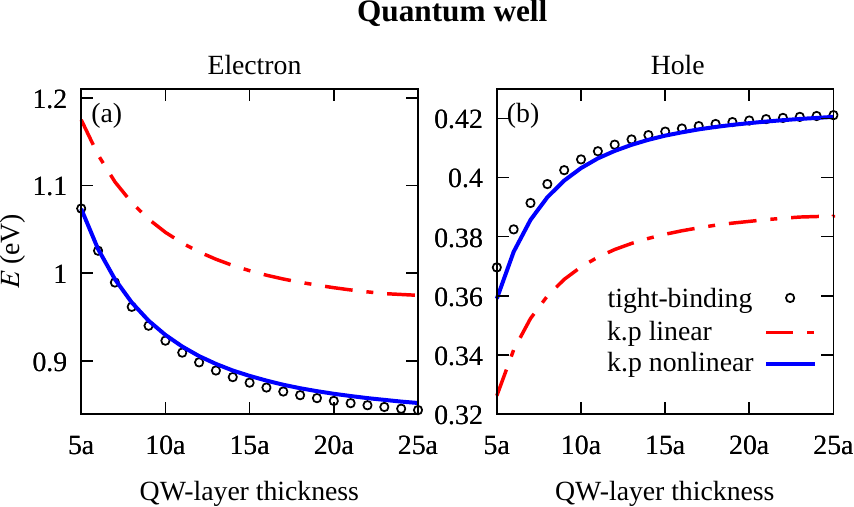}
		\end{center}
		\caption{\label{fig:qw}\textcolor{gray}(Color online) Energy of the electron (a) and the hole (b) ground state, as a function of the QW width. Energy $E = 0$ corresponds to the (unstrained)
			GaAs valence-band edge. For the linear strain approximation we took the deformation potentials from Ref.\cite{vurgaftman01} (see Table~\ref{tab:dps}). }
	\end{figure}
	\subsection{Quantum Wells}
	In the previous section, we found that the non-linear, 8-band $\kp$ model is capable to reproduce accurately bulk band edges evolution under strain, as given by the atomistic tight-binding model. In the following, we will verify how both methods compare for nanostructures, where the quantum confinement plays a significant role.
	We start by calculating the ground state energy of the electron [Fig.\ref{fig:qw}(a)] and hole [Fig.\ref{fig:qw}(b)], in a quasi-two dimensional system, namely the InAs quantum well embedded in bulk GaAs. We study electron and hole ground state energies as a function of quantum well width changing from $5a$ to $25a$, where $a$ is the lattice constant. This corresponds to quantum well thickness from approximately $3$ to $15$~nm. 
	The computational domain size for the strain calculation is taken $220a \times 220a \times 220a$, while the single particle states are calculated in a $80a \times 80a \times 60a$ box.~\cite{zielinski-multiscale}
	Fig.~\ref{fig:qw} shows results obtained by the tight binding calculation, the standard Bir-Pikus model, and the Bir-Pikus model with non-linear terms. Apparently (and similarly to the strained bulk results) the linear Bir-Pikus model significantly overestimates the single particle electron energy and as well it underestimates the ground hole state energy  (please note reversed ordering of hole states as compared to the electron).
	This is however expected since the quantum well is under high biaxial strain (the lattice mismatch between GaAs and InAs is about $7$\%), which according to Fig.~\ref{fig:edges}(c) has strongly non-linear character and is not well reproduced by linear $\kp$ approach already at the bulk level. 
	In particular, in case of linear $\kp$ the electron ground state energy $e_1$ is overestimated by approximately $100$~meV with respect to the tight-binding approach. 
	Additionally, approximately $30$~meV difference is also present in the energy of the hole ground state $h_1$. Overall, the linear $\kp$ systematically overestimates the single particle energy gap $e_1-h_1$ by approximately $130$~meV.
	On the other hand, we obtain an excellent agreement between the 8-band $\kp$ and the tight-binding method, if the terms second order in strain are taken into account.
	For all considered quantum well thicknesses we obtain at most $\approx 10$~meV difference between both methods, with only some variations due to different quantum well heights.
	
	\subsection{Quantum Dots}
	\subsubsection*{Disk-shaped quantum dots}
	\begin{figure}[!tb]
		\begin{center}
			\includegraphics[width=3.5in]{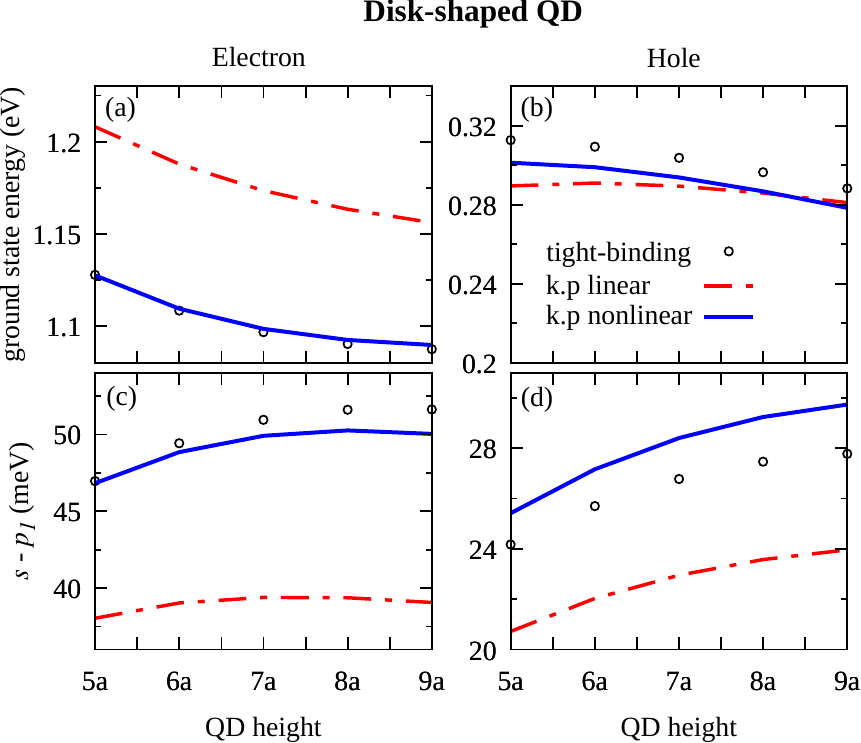}
		\end{center}
		\caption{\label{fig:disk}\textcolor{gray}({Color online) Energy of the electron and hole ground states, as a function of the QD height, where the base radius is fixed as a $21a$. The width of the wetting layer is $a$. } }
	\end{figure}    
	Next, we have calculated several lowest electron and hole states for a series of disk-shaped InAs/GaAs quantum dots. Here, we have assumed fixed quantum dot basis diameter equal to approximately $25$~nm and systematically varied quantum dot height from approximately $3$ to $5.4$~nm ($5$ to $9$~$a$).
	The dot is placed on a lattice constant thick ($\approx 0.6$~nm) wetting layer.
	Similarly to results for quantum wells, for disk-shaped quantum dots, if the second order strain terms are taken into account, we obtain an excellent agreement for the electron ground state [Fig.\ref{fig:disk}(a)]. 
	Additionally, we study here the energy difference [Fig.\ref{fig:disk}(c)] between the first excited electron and the ground electron states (i.e. energy difference between electron $p-$ and $s-$ shells, i.e. $e_2-e_1$). 
	This energy difference increases with the quantum dot height, and the magnitude of this spacing is also much better (within few meV error) reproduced by the augmented, non-linear $\kp$ approach as compared to the straightforward linear $\kp$, with predictions systematically smaller by approximately $10$~meV. 
	
	In the case of the hole ground state energy [Figs.\ref{fig:disk}(b)], the relative discrepancies between 8-band $\kp$ and tight-binding model become somewhat larger. This could be expected since Jancu's~\cite{Jancu1998} tight-binding model is a 20-band, atomistic model with d-orbitals, and hole states in quantum dots have complicated multi-band character. However, interestingly all three models predict quite similar energy of the ground hole state ($\approx 0.3$~eV), with non-linear results only several meV's closer to the tight-binding results for flat (small height) quantum dots, than those given by linear $\kp$. 
	Moreover, contrarily to a quantum well, the ground hole state energy of a disc-shape quantum dot rather weakly depends on quantum dot height. 
	Actually, it even shows an opposite trend with respect to the height, due to strain distribution changing its character with an increasing quantum dot height.~\cite{zielinski-tallqd} Therefore, size-dependent strain distribution in finite size quantum dots (rather than in quasi-2D quantum well) seems to play here a dominant role. Should strain effects be artificially neglected (see the Appendix B) the ground hole state energy would simply increase with dot's height due to reduced confinement. In fact, as shown in the Appendix, $\kp$ and tight-binding results for disk-shaped quantum dots with strain artificially neglected are quite similar, again demonstrating the key role of strain in modeling of quantum dots, and in particular the role of non-linear strain treatment in $\kp$.
	
	To summarize, the single particle gap ($e_1-h_1$) in disk-shaped InAs/GaAs quantum dots is synthetically overestimated by $\approx 100$~meV by the linear $\kp$ approach (mostly due to large overestimation of $e_1$), and only by $\approx 15$~meV by the non-linear $\kp$ method (mostly due to small underestimation of $h_1$).
	
	The energy difference between the ground and the first excited hole states ($s-$ and $p-$ shells spacing, i.e., $h_1-h_2$,\footnote{Note reversed ordering of hole states with respect to the electron}) increases with the quantum dot height as in the case of electron [Figs.\ref{fig:disk}(d)], again with non-linear $\kp$ predictions closer to the tight-binding, than the linear approach results. The second order $\kp$ overestimates the magnitude of this spacing by $\approx 2$~meV, whereas the linear $\kp$ underestimates this difference by $\approx4$~meV. Therefore, both for electron and hole $s-p$ shell spacings, the agreement between the non-linear $\kp$ and the tight-binding could be described as satisfactory considering how inherently different (continuous media vs atomistic) $\kp$ and tight-binding models are. 
	
	\subsubsection*{Lens-shaped quantum dots}
	\begin{figure}[!tb]
		\begin{center}
			\includegraphics[width=3.5in]{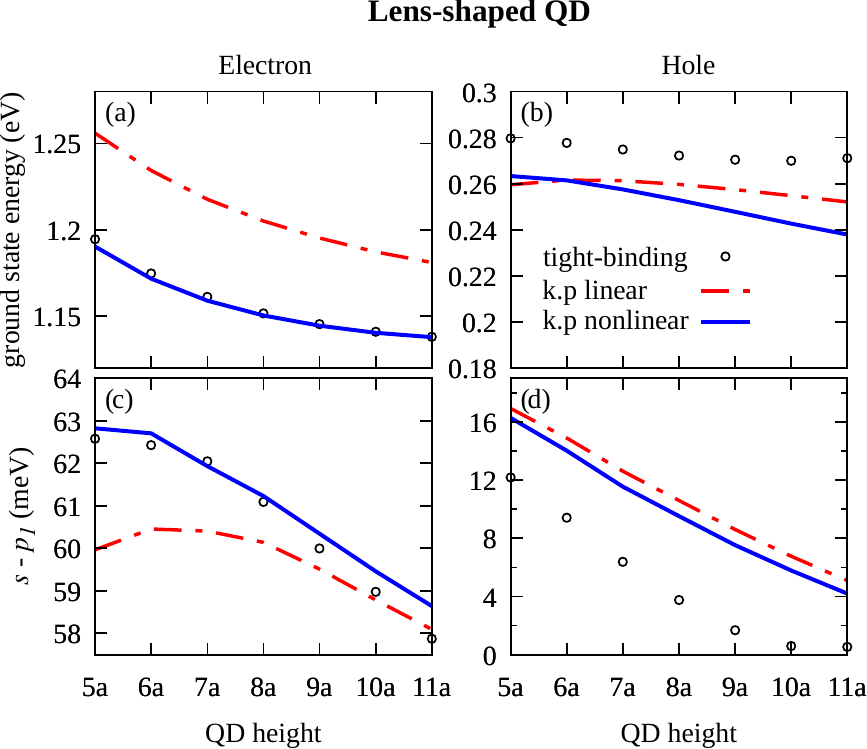}
		\end{center}
		\caption{\label{fig:lens}\textcolor{gray}({Color online) Energy of the electron and hole ground states, as a function of the QD height, where the base radius is fixed as a $20a$. The width of the wetting layer is $a$. } }
	\end{figure}
	It is instructive to verify, how non-linear $\kp$ performs for other geometries, rather than flat quantum wells and disk-shaped quantum dots.
	Therefore, Fig.\ref{fig:lens} shows results obtained, for lens-shaped quantum dot, with its height varying from approximately $3$ to $\approx6.6$~nm  ($5$ to $11$~$a$), while keeping the base diameter fixed to about $24$~nm. 
	Similarly to the previous case, the lens-shaped quantum dot is placed on a lattice constant thick ($\approx 0.6$~nm) wetting layer.
	Again, for the ground electron state we obtain an excellent agreement [Fig.\ref{fig:lens}(a)] between the non-linear $\kp$ and the tight-binding.
	
	Again, a very good agreement [Fig.\ref{fig:lens}(b)] is found for $p-s$ electron energy spacing (i.e., $e_2-e_1$), with the non-linear method differing by at most $1$~meV from the tight-binding.
	Notably, for the lens-shaped quantum dot (and differently from the disk-shaped quantum dots) the electron $p-s$ energy difference is reduced with the quantum dot height.

	Regarding, the ground hole state [Fig.\ref{fig:lens}(b)], similarly to the disk-shape quantum dots, both $\kp$ variants underestimate this energy by about $20$ to $30$~meV. Thus, the single particle band gap ($e_1-h_1$) is overestimated by $\approx 70$~meV by the linear approach, and only by at most $\approx 30$~meV by the non-linear approach. 
	Again, should the strain effect be artificially neglected (see the Appendix B), the difference between 8-band $\kp$ and the tight-binding predicted ground hole energies would be withing several meV's.
	
	A more notable difference between various methods is observed for the $s-p$ energy spacing of hole states ($h_2-h_1$) which is systematically overestimated ($\approx 4$~meV) by both $\kp$ approaches. The difference between atomistic and continuous media approximation is thus somewhat larger for hole states in lens-shaped quantum dot than in disk-shaped quantum dots. This difference is present for both unstrained (the Appendix) and strained quantum dots, yet strain further increases its magnitude.
	
	Part of this discrepancy is probably related to shear strains which are more notable in lens-shaped quantum dots due to their curved shape. 
	Deformation potentials due to shear strains are in fact reproduced with limited accuracy by Jancu's tight-binding parameterization.~\cite{Jancu1998,Jancu2007}
	Having said the above, we should note that is not our goal to claim that the continuous media approximation (with complicated atoms to grid strain transfer) in all considered cases is able reproduce the atomistic results. 

	\section{Conclusions}
	We have investigated the influence of strain nonlinearity on the electron and hole energy levels in semiconductor nanostructures. We have used Bir-Pikus Hamiltonian extended by the terms which are quadratic in strain. We have obtained the set of parameters for InAs and GaAs bulk semiconductors by fitting to the results obtained from the $sp^3d^5s^*$ tight-binding method. 
	Next, we have provided the ready-to-use parameterization for the effective term used to improve the accuracy of the standard effective mass method.
	Then, we have calculated the electron and hole states for quantum wells, and various quantum dot systems. 
	We have shown that, while the standard (linear in strain) $8$-band $\kp$ model overestimates the electron and underestimates the hole energy, a very good agreement to the tight-binding results can be achieved if the terms second order in strain are taken into account. 
	This agreement is particularly good for electron states confined in quantum wells and flat quantum dot systems. 
	More discrepancies are found for hole states, especially in curved, lens-shaped quantum dots.
	We are convinced this differences origin from the different treatment of strains (in particular shear strain) by $\kp$ and tight-binding, with a further research needed to tract these divergences.
	
	\label{sec:concl}
	
	\acknowledgments
	K.G. acknowledges the support from the National Centre for Research and 
	Development in Poland within the FI-SEQUR Project of the 2nd 
	Poland-Berlin Photonics Programme, grant No. 2/POLBER-2/2016 (project 
	value 2 089 498 PLN) and from the Polish National Agency for Academic Exchange. M.Z. acknowledges the support from the Polish National Science Centre based on Decision No. 2018/31/B/ST3/01415.
	Calculations have been carried out using
	resources provided by Wroclaw Centre
	for Networking and Supercomputing (\url
	{http://wcss.pl}), Grant No.~203.

	\appendix
	\section{Second order strain Hamiltonian}
	
	According to the standard $\kp$ model, the Hamiltonian can be written by\cite{Winkler2003}
	\begin{align*}
	H_{\kk} &= \frac{p^2}{2 m_0} + \frac{\hbar^2 k^2}{2 m_0} + \frac{\hbar}{m_0} \kk \cdot \bm{p}  + V_0(\rr) \\
	& + \frac{\hbar}{4m^2_0 c^2} ( \bm{p} + \hbar \kk  ) \cdot \bm{\sigma} \times (\grad{V_0(\rr)} ),
	\end{align*}
	where $m_0$ is the free electron mass, $V_0(\rr)$ describes the crystal potential, $c$ is the speed of light, and
	$\bm{\sigma} = (\sigma_x,\sigma_y,\sigma_z)$ denotes vector of Pauli matrices.
	
	The strain changes crystal lattice and lowers symmetry of the system. The Hamiltonian of the deformed crystal $H'_{\kk'}$  can be described in terms of the transformed coordinates $\rr'$ \cite{Bir1974}.
	Then, the Hamiltonian is written in basis of $\kk=0$ Bloch functions $\ket{\mu \lambda}$, where  $\mu$ describes its orbital part and $\lambda$ is a spin\cite{Winkler2003}. The matrix elements, expressed up to the second order in strain tensor elements, takes the form
	\begin{align*}
	\bra{ \lambda \nu} H'_0  \ket{\mu \lambda'} &= E_{\nu} \delta_{\nu \mu} \delta_{\lambda \lambda'} + \sum_{ij} D^{\nu\mu}_{ij} \eps_{ij} \delta_{\lambda \lambda'} \\ &   + \sum_{ijkl} F^{\nu\mu}_{ijkl} \eps_{ij} \eps_{kl} \delta_{\lambda \lambda'},
	\end{align*}
	where $E_{\nu}$ describes the unstrained band edges, $D^{\nu\mu}_{ij}$ and $F^{\nu\mu}_{ijkl}$ are first- and second-order deformation potentials respectively.
	
	With the group theory, one can predict the number of independent tensor compontents ($N_0$) for a given symmetry point group. If tensor $\bm S$ connects $\bm A$  and $\bm B$ (eg. $A_i = \sum_j S_{ij} B_j$), and  $\bm A$, $\bm B$ belong to the $\mathcal{D}_A$, $\mathcal{D}_B$ representations respectively, the number of independent $\bm S$ components is given by the number of trivial representations in \cite{Bir1974}
	$$
	\mathcal{D}_S = \mathcal{D}_{A} \times \mathcal{D}^*_{B}.
	$$
	Let us consider the tensor describing deformation potentials ($\hat D$) as $\bm{S}$, a given block of the Hamiltonian as $\bm{A}$, and the strain tensor as $\bm{B}$. The strain tensor $\hat \eps$ belongs to the representation $\Gamma_5 \otimes \Gamma_5 = \Gamma_1 \oplus \Gamma_3 \oplus \Gamma_4 \oplus \Gamma_5$, and since it is symmetric, the part related to $\Gamma_4$ vanishes.\cite{Bir1974,Yu2010} If we neglect the spin-orbit coupling, the conduction band (cb.) is related to the $\Gamma_1 \otimes \Gamma_1$, and the valence band (vb.) to the $\Gamma_5 \otimes \Gamma_5$ representations. Hence, in the case of the $D^{\nu\mu}_{ij}$, the number of the independent components is: one for the cb., and three for the vb. block. In consequence, only four deformation potentials are needed: $a_\mathrm{c}$, $a_\mathrm{v}$, $b_\mathrm{v}$, and $d_\mathrm{v}$ (we do not consider here, the block connecting cb. with vb.).
	
	In the case the strain quadratic terms, the product $\hat \epsilon \otimes \hat \epsilon$ belongs to  $ (\Gamma_5 \otimes \Gamma_5) \otimes (\Gamma_5 \otimes \Gamma_5) = 3 \Gamma_1 \oplus 3 \Gamma_3 \oplus 3\Gamma_5$, which leads to $3$ independent terms in the cb. block ($a^{\mathrm{(2a)}}_\mathrm{c}$, $a^{\mathrm{(2b)}}_\mathrm{c}$, $a^{\mathrm{(2c)}}_\mathrm{c}$) and $9$ components in the vb. block (which we denote as $a^{\mathrm{(2a)}}_\mathrm{v}, a^{\mathrm{(2b)}}_\mathrm{v}, a^{\mathrm{(2c)}}_\mathrm{v}, b^{\mathrm{(2a)}}_\mathrm{v}, b^{\mathrm{(2b)}}_\mathrm{v}, b^{\mathrm{(2c)}}_\mathrm{v}, d^{\mathrm{(2a)}}_\mathrm{v}, d^{\mathrm{(2b)}}_\mathrm{v}, d^{\mathrm{(2c)}}_\mathrm{v}$).  In presence of spin-orbit interaction, the Hamiltonian blocks are described in terms of the double group representation, but it does not change the number of independent tensor components related to strain.
	
	We write the Hamiltonian using the invariant expansion method\cite{Luttinger1955}. We note, that strain tensor elements transform like symmetrized products  $\{k_i,k_j\}$. Then,we utilize the list of irreducible tensor components for the point group $T_d$, given (up to the 4th order in $k$) in the Appendix C of Ref.\cite{Winkler2003}. This leads to the form
	\begingroup
	\allowdisplaybreaks
	\begin{subequations}
		\begin{align*}
		H^{\mathrm{(2)}}_{\mathrm{str},\mathrm{6c6c}} &=  a^{\mathrm{\mathrm{(2a)}}}_{\mathrm{c}} (\exx + \eyy + \ezz)^{2} + a^{\mathrm{(2b)}}_{\mathrm{c}} (\exx \eyy + \cp)\\
		&\phantom{=} + a^{\mathrm{(2c)}}_{\mathrm{c}} (\exy^2 + \cp),\\
		H^{\mathrm{(2)}}_{\mathrm{str},\mathrm{8v8v}} &=  a^{\mathrm{\mathrm{(2a)}}}_{\mathrm{v}} (\exx + \eyy + \ezz)^{2} + a^{\mathrm{(2b)}}_{\mathrm{v}} (\exx \eyy + \cp)\\
		&\phantom{=} + a^{\mathrm{(2c)}}_{\mathrm{v}} (\exy^2 + \cp)\\
		&\phantom{=} - b^{\mathrm{(2a)}}_{\mathrm{v}} \left[ \left( J^{2}_{x}  - \frac{1}{3} J^{2} \right)\epsilon^{2}_{xx} + \cp  \right] \\
		&\phantom{=} - b^{\mathrm{(2b)}}_{\mathrm{v}} \left[ \left( J^{2}_{x}  - \frac{1}{3} J^{2} \right)\eyy\ezz + \cp  \right],\\
		&\phantom{=} - b^{\mathrm{(2c)}}_{\mathrm{v}} \left[ \left( J^{2}_{x}  - \frac{1}{3} J^{2} \right)\eyz^2 + \cp  \right],\\
		& \phantom{=} - \frac{2}{\sqrt{3}} d^{\mathrm{(2a)}}_{\mathrm{v}} \left[ \{J_{x},J_{y}\} \exy \ezz  + \cp \right],\\
		& \phantom{=} - \frac{2}{\sqrt{3}} d^{\mathrm{(2b)}}_{\mathrm{v}} \left[ \{J_{x},J_{y}\} \ezx \eyz  + \cp \right],\\
		& \phantom{=} - \frac{2}{\sqrt{3}} d^{\mathrm{(2c)}}_{\mathrm{v}} \left[ \{J_{x},J_{y}\} (\exx + \eyy) \exy  + \cp \right],\\
		H^{\mathrm{(2)}}_{\mathrm{str},\mathrm{8v7v}} &=  -3 b^{\mathrm{(2a)}}_{\mathrm{v}} \left( U_{xx} \exx^2 + \cp \right) \\
		&\phantom{=} -3 b^{\mathrm{(2b)}}_{\mathrm{v}} \left( U_{xx} \eyy \ezz + \cp \right) \\
		&\phantom{=} -3 b^{\mathrm{(2c)}}_{\mathrm{v}} \left( U_{xx} \eyz^2 + \cp \right) \\
		&\phantom{=} -\sqrt{3} d^{\mathrm{(2a)}}_{\mathrm{v}} \left( 2 U_{xy} \exy \ezz + \cp \right) \\
		&\phantom{=} -\sqrt{3} d^{\mathrm{(2b)}}_{\mathrm{v}} \left( 2 U_{xy} \ezx \eyz + \cp \right) \\
		&\phantom{=} -\sqrt{3} d^{\mathrm{(2c)}}_{\mathrm{v}} \left[ 2 U_{xy} (\exx + \eyy) \exy + \cp \right], \\
		H^{\mathrm{(2)}}_{\mathrm{str},\mathrm{7v7v}} &=  a^{\mathrm{\mathrm{(2a)}}}_{\mathrm{v}} (\exx + \eyy + \ezz)^{2} + a^{\mathrm{(2b)}}_{\mathrm{v}} (\exx \eyy + \cp)\\
		&\phantom{=} + a^{\mathrm{(2c)}}_{\mathrm{v}} (\exy^2 + \cp).
		\end{align*}
	\end{subequations}
	\endgroup
	We neglected quadratic strain in off-diagonal blocks between the cb. and valence bands (namely $H^{\mathrm{(2)}}_{\mathrm{str},\mathrm{6c8v}}$ and $H^{\mathrm{(2)}}_{\mathrm{str},\mathrm{6c7v}}$).
	Due to large number of free parameters and uncertainty in the value of Kleinman parameter ($\zeta$), we skipped the fitting of parameters describing the shear strain ($a^{\mathrm{(2c)}}_\mathrm{c}$, $a^{\mathrm{(2c)}}_\mathrm{v}$, $b^{\mathrm{(2c)}}_\mathrm{v}$, $d^{\mathrm{(2a)}}_\mathrm{v}$, $d^{\mathrm{(2b)}}_\mathrm{v}$, and $d^{\mathrm{(2c)}}_\mathrm{v}$). We also neglected terms containing products of wave vector and strain tensor elements.
	
	\section{Quantum dots with strain effects neglected}
	\begin{figure}[!tb]
		\begin{center}
			\includegraphics[width=3.5in]{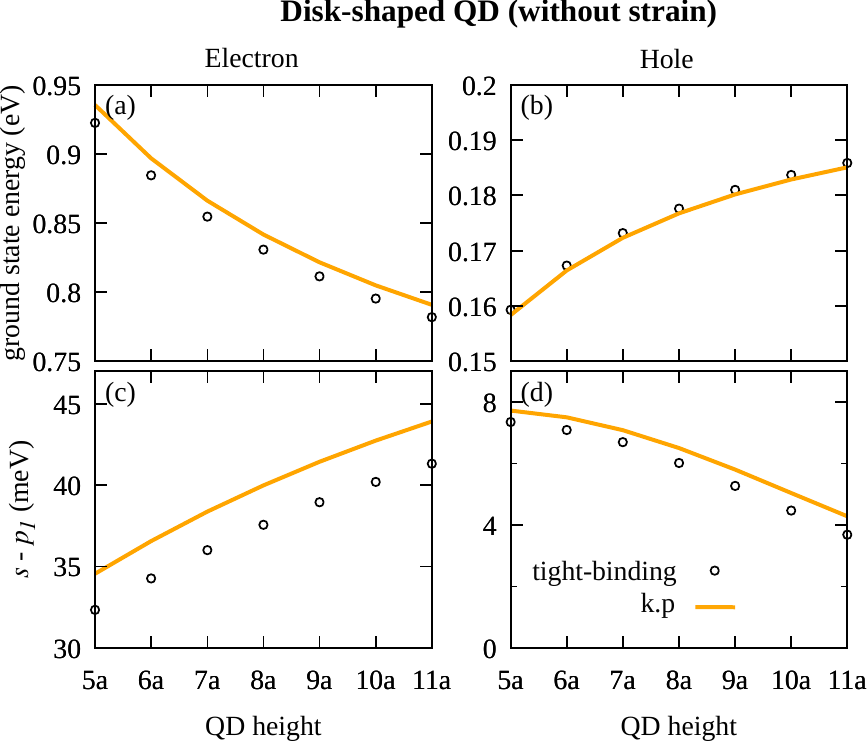}
		\end{center}
		\caption{\label{fig:nostraindisk}\textcolor{gray}({Color online) Energy of the electron and hole ground states, as a function of the QD height, where the base radius is fixed as a $21a$. The width of the wetting layer is $a$. } }
	\end{figure}
	\begin{figure}[!tb]
		\begin{center}
			\includegraphics[width=3.5in]{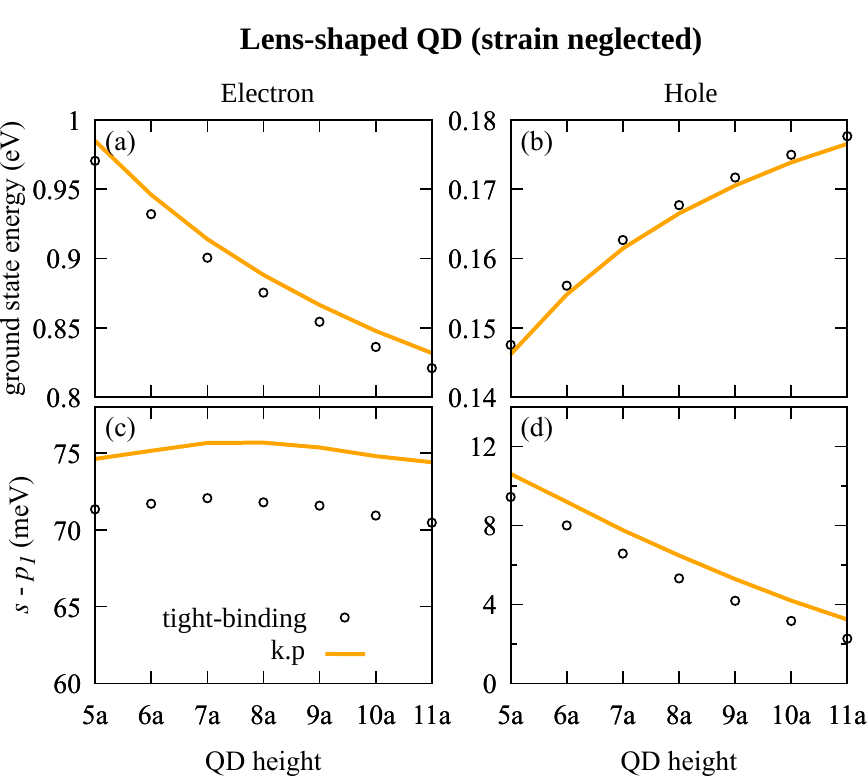}
		\end{center}
		\caption{\label{fig:nostrainlens}\textcolor{gray}({Color online) Energy of the electron and hole ground states, as a function of the QD height, where the base radius is fixed as a $20a$. The width of the wetting layer is $a$. } }
	\end{figure}
	Here, for completeness, we show results obtained for disk-shaped [Fig.~\ref{fig:nostraindisk}] and lens-shaped [Fig.~\ref{fig:nostrainlens}] quantum dots of the same shapes, dimensions, and compositions, as in the main text, yet with strain effects artificially neglected (i.e. assuming identical bond lengths of InAs and GaAs, as well as, perfect tetrahedral bond angles).
	
	Both figures show the same trends of decreasing (increasing) ground electron (hole) state energy with quantum dot height, consistent with the quantum confinement effect.
	For both geometries the agreement between $\kp$ and the tight-binding description of carriers ground state energies is very good.
	Somewhat smaller agreement is found for the energy difference between the ground and the first excited states. Here, the $\kp$ results systematically overestimate the magnitude of this splitting, with respect to the tight-binding, for both disk- and lens-shaped quantum dots, and both for the electron and the hole.
		
	In the unstrained case, the hole $s-p$ shells spacing, ($h_1-h_2$ energy spacing) [Fig.~\ref{fig:nostrainlens} (c)] is reduced with growing quantum dot height for both disk- and lens-shaped systems.
	However, interestingly, the electron $s-p$ spacing either increases with the quantum dot height for disk-shaped nanostructures or show a rather flat (and non-monotonous) trend for lens-shaped systems.
	In all considered cases, contrary to a situation where the strain is accounted for, $\kp$ and tight-binding results are both qualitatively and quantitatively similar.
	
	\bibliographystyle{prsty}
	\bibliography{abbr,../library.bib,michal.bib}

\begin{thebibliography}{10}

\bibitem{delerue-book}
C. Delerue and M. Lannoo, {\em Nanostrucutres: Theory and Modelling} (Springer,
  New York, 2004).

\bibitem{bimberg-book}
D. Bimberg, M. Grundmann, and N. Ledentsov, {\em Quantum Dot Heterostructures}
  (Wiley, Chichester, 1999).

\bibitem{michler}
{\em Topics in Applied Physics}, edited by P. Michler (Springer, New York,
  2003), Vol.~90.

\bibitem{michler2017quantum}
P. Michler, {\em {Quantum Dots for Quantum Information Technologies}}
  (Springer, 2017), Vol.~237.

\bibitem{tartakovskii_2012}
A. Tartakovskii, {\em {Quantum Dots: Optics, Electron Transport and Future
  Applications}} (Cambridge University Press, 2012).

\bibitem{bjork}
M.~T. Bjork, C. Thelander, A. Hansen, L. Jensen, M. Larsson, L.~R. Wallenberg,
  and L. Samuelson, Nano Lett. {\bf 4},  1621  (2004).

\bibitem{borgstrom-nwqd}
M.~T. Borgström, V. Zwiller, E. Müller, and A. Imamoglu, Nano Lett. {\bf 5},
  1439  (2005).

\bibitem{dery2004nature}
H. Dery, E. Benisty, A. Epstein, R. Alizon, V. Mikhelashvili, G. Eisenstein, R.
  Schwertberger, D. Gold, J. Reithmaier, and A. Forchel, J. Appl. Phys. {\bf
  95},  6103  (2004).

\bibitem{Khan2014}
M.~Z.~M. Khan, T.~K. Ng, and B.~S. Ooi, Prog. Quantum Electron. {\bf 38},  237
  (2014).

\bibitem{vls-wagner}
R.~S. Wagner and W.~C. Ellis, Appl. Phys. Lett. {\bf 4},  89  (1964).

\bibitem{dalacu-selective}
D. Dalacu, A. Kam, D.~G. Austing, X. Wu, J. Lapointe, G.~C. Aers, and P.~J.
  Poole, Nanotechnology {\bf 20},  395602  (2009).

\bibitem{Zwanenburg_RMP13}
F.~A. Zwanenburg, A.~S. Dzurak, A. Morello, M.~Y. Simmons, L.~C.~L. Hollenberg,
  G. Klimeck, S. Rogge, S.~N. Coppersmith, and M.~A. Eriksson, Rev. Mod. Phys.
  {\bf 85},  961  (2013).

\bibitem{Usman2016}
M. Usman, J. Bocquel, J. Salfi, B. Voisin, A. Tankasala, R. Rahman, M.~Y.
  Simmons, S. Rogge, and L.~C.~L. Hollenberg, Nat. Nanotechnol. {\bf 11},  763
  EP   (2016).

\bibitem{szabo-ostlund}
A. Szabo and N. Ostlund, {\em Modern Quantum Chemistry} (MacMillan, 1983).

\bibitem{siesta}
J.~M. Soler, E. Artacho, J.~D. Gale, A. García, J. Junquera, P. Ordejón, and
  D. Sánchez-Portal, J. Phys Cond. Matt. {\bf 14},  2745  (2002).

\bibitem{stier-grundmann-bimberg}
O. Stier, M. Grundmann, and D. Bimberg, Phys. Rev. B {\bf 59},  5688  (1999).

\bibitem{kp-schliwa}
A. Schliwa, M. Winkelnkemper, and D. Bimberg, Phys. Rev. B {\bf 76},  205324
  (2007).

\bibitem{Schliwa111}
A. Schliwa, M. Winkelnkemper, A. Lochmann, E. Stock, and D. Bimberg, Phys. Rev.
  B {\bf 80},  161307  (2009).

\bibitem{Schulz111}
S. Schulz, M.~A. Caro, E.~P. O'Reilly, and O. Marquardt, Phys. Rev. B {\bf 84},
   125312  (2011).

\bibitem{kp14}
S. Tomić and N. Vukmirović, J. Appl. Phys. {\bf 110},    (2011).

\bibitem{pryor-zunger}
C. Pryor, J. Kim, L.~W. Wang, A.~J. Williamson, and A. Zunger, J. Appl. Phys.
  {\bf 83},  2548  (1998).

\bibitem{wojs}
A. Wojs, P. Hawrylak, S. Fafard, and L. Jacak, Phys. Rev. B {\bf 54},  5604
  (1996).

\bibitem{weidong-localization}
W. Sheng and J.-P. Leburton, Appl. Phys. Lett. {\bf 81},  4449  (2002).

\bibitem{sheng-cheng-prb2005}
W. Sheng, S.-J. Cheng, and P. Hawrylak, Phys. Rev. B {\bf 71},  035316  (2005).

\bibitem{ardelt16}
P.-L. Ardelt, K. Gawarecki, K. M{\"{u}}ller, A. Waeber, A. Bechtold, K.
  Oberhofer, J. Daniels, F. Klotz, M. Bichler, T. Kuhn, H. Krenner, P.
  Machnikowski, and J. Finley, Phys. Rev. Lett. {\bf 116},  77401  (2016).

\bibitem{climente08}
J.~I. Climente, M. Korkusinski, G. Goldoni, and P. Hawrylak, Phys. Rev. B {\bf
  78},  115323  (2008).

\bibitem{leger}
Y. L\'eger, L. Besombes, L. Maingault, and H. Mariette, Phys. Rev. B {\bf 76},
  045331  (2007).

\bibitem{marynski2013electronic}
A. Mary{\'{n}}ski, G. S{\c{e}}k, A. Musia{\l}, J. Andrzejewski, J. Misiewicz,
  C. Gilfert, J.~P. Reithmaier, A. Capua, O. Karni, D. Gready, and Others, J.
  Appl. Phys. {\bf 114},  94306  (2013).

\bibitem{gawelczyk2017exciton}
M. Gawe{\l}czyk, M. Syperek, A. Mary{\'{n}}ski, P. Mrowi{\'{n}}ski, K.
  Gawarecki, J. Misiewicz, A. Somers, J.~P. Reithmaier, S. H{\"{o}}fling, G.
  S{\c{e}}k, and Others, Phys. Rev. B {\bf 96},  245425  (2017).

\bibitem{andlauer09}
T. Andlauer and P. Vogl, Phys. Rev. B {\bf 79},  45307  (2009).

\bibitem{Gawarecki2018a}
K. Gawarecki, Phys. Rev. B {\bf 97},  235408  (2018).

\bibitem{Gawelczyk2018a}
M. {Gawe{\l}czyk Micha{\l}and Krzykowski}, K. Gawarecki, and P. Machnikowski,
  Phys. Rev. B {\bf 98},  75403  (2018).

\bibitem{Note1}
We note that there are some intermediate schemes in particular the EBOM method,
  which however does not have atomistic spatial resolution.

\bibitem{Zunger-EPM2}
A.~J. Williamson, L.~W. Wang, and A. Zunger, Phys. Rev. B {\bf 62},  12963
  (2000).

\bibitem{bester-zunger-nair}
G. Bester, S. Nair, and A. Zunger, Phys. Rev. B {\bf 67},  161306  (2003).

\bibitem{bester-zunger-p-shell}
G. Bester and A. Zunger, Phys. Rev. B {\bf 71},  045318  (2005).

\bibitem{singh-bester-ordering}
R. Singh and G. Bester, Phys. Rev. B {\bf 84},  241402  (2011).

\bibitem{singh-bester-lower}
R. Singh and G. Bester, Phys. Rev. Lett. {\bf 104},  196803  (2010).

\bibitem{slater-koster}
J.~C. Slater and G.~F. Koster, Phys. Rev. {\bf 94},  1498  (1954).

\bibitem{vogl}
P. Vogl, H.~P. Hjalmarson, and J.~D. Dow, J. Phys. Chem. Solids {\bf 44},  365
   (1983).

\bibitem{jancu}
J.-M. Jancu, R. Scholz, F. Beltram, and F. Bassani, Phys. Rev. B {\bf 57},
  6493  (1998).

\bibitem{santoprete03}
R. Santoprete, B. Koiller, R.~B. Capaz, P. Kratzer, Q.~K.~K. Liu, and M.
  Scheffler, Phys. Rev. B {\bf 68},  235311  (2003).

\bibitem{saito-arakawa}
T. Saito and Y. Arakawa, Physica E {\bf 15},  169  (2002).

\bibitem{schultz12}
S. Schulz, M.~A. Caro, E.~P. O'Reilly, and O. Marquardt, Phys. Status Solidi
  {\bf 249},  521  (2012).

\bibitem{schulz-schum-czycholl}
S. Schulz, S. Schumacher, and G. Czycholl, Phys. Rev. B {\bf 73},  245327
  (2006).

\bibitem{alloyed-qd}
G. Klimeck, F. Oyafuso, T. Boykin, R. Bowen, and P. von Allmen, Comp. Modeling
  in Eng. and Sci. (CMES) {\bf 3},  601  (2002).

\bibitem{boykin-fit}
G. Klimeck, R.~C. Bowen, T.~B. Boykin, and T.~A. Cwik, Superlattices and
  Microstructures {\bf 27},  519   (2000).

\bibitem{voisin-fit}
R. Benchamekh, F. Raouafi, J. Even, F. {Ben Cheikh Larbi}, P. Voisin, and J.-M.
  Jancu, Phys. Rev. B {\bf 91},  45118  (2015).

\bibitem{jaskolski06}
W. Jaskolski, M. Zielinski, G.~W. Bryant, J. Aizpurua, W. Jask{\'{o}}lski, M.
  Zieli{\'{n}}ski, G.~W. Bryant, and J. Aizpurua, Phys. Rev. B {\bf 74},
  195339  (2006).

\bibitem{jaskolski04}
W. Jask{\'{o}}lski and M. Zieli{\'{n}}ski, Acta Phys. Pol. A {\bf 106},  193
  (2004).

\bibitem{zielinski12}
M. Zieli{\'{n}}ski, Phys. Rev. B {\bf 86},  115424  (2012).

\bibitem{Zielinski2013}
M. Zieli{\'{n}}ski, J. Phys. Condens. Matter {\bf 25},  465301  (2013).

\bibitem{bayer-eh}
M. Bayer, G. Ortner, O. Stern, A. Kuther, A.~A. Gorbunov, A. Forchel, P.
  Hawrylak, S. Fafard, K. Hinzer, T.~L. Reinecke, S.~N. Walck, J.~P.
  Reithmaier, F. Klopf, and F. Sch\"afer, Phys. Rev. B {\bf 65},  195315
  (2002).

\bibitem{takagahara}
T. Takagahara, Phys. Rev. B {\bf 62},  16840  (2000).

\bibitem{Swiderski2017}
M. {\'{S}}widerski and M. Zieli{\'{n}}ski, Phys. Rev. B {\bf 95},  125407
  (2017).

\bibitem{rand-mlinar-zunger}
V. Mlinar and A. Zunger, Phys. Rev. B {\bf 79},  115416  (2009).

\bibitem{Zielinski-natural}
M. Zieliński, K. Gołasa, M.~R. Molas, M. Goryca, T. Kazimierczuk, T.
  Smoleński, A. Golnik, P. Kossacki, A.~A.~L. Nicolet, M. Potemski, Z.~R.
  Wasilewski, and A. Babiński, Phys. Rev. B {\bf 91},  085303  (2015).

\bibitem{dash-mrowinski}
P. Mrowinski, M. Zieliński, M. Świderski, J. Misiewicz, A. Somers, J.~P.
  Reithmaier, S. Höfling, and G. Sęk, Phys. Rev. B {\bf 94},  115434  (2016).

\bibitem{bryant-prl}
G.~W. Bryant, M. Zieliński, N. Malkova, J. Sims, W. Jask\'{o}lski, and J.
  Aizpurua, Phys. Rev. Lett. {\bf 105},  067404  (2010).

\bibitem{bryant-prb}
G.~W. Bryant, M. Zieliński, N. Malkova, J. Sims, W. Jaskólski, and J.
  Aizpurua, Phys. Rev. B {\bf 84},  235412  (2011).

\bibitem{Zielinski.PRB.2015}
M. Zieliński, Y. Don, and D. Gershoni, Phys. Rev. B {\bf 91},  085403  (2015).

\bibitem{wojnar2014strain}
P. Wojnar, M. Zielinski, E. Janik, W. Zaleszczyk, T. Wojciechowski, R. Wojnar,
  M. Szymura, {\L}. K{\l}opotowski, L. Baczewski, A. Pietruchik, {\it et~al.},
  Appl. Phys. Lett. {\bf 104},  163111  (2014).

\bibitem{Diaz.PhysRevB.75.245433}
J.~G. D{\'{i}}az, G.~W. Bryant, W. Jask{\'{o}}lski, and M. Zieli{\'{n}}ski,
  Phys. Rev. B {\bf 75},  245433  (2007).

\bibitem{zhang}
L. Zhang, J.-W. Luo, A. Zunger, N. Akopian, V. Zwiller, and J.-C. Harmand, Nano
  Lett. {\bf 10},  4055  (2010).

\bibitem{nika-cpqd}
N. Akopian, G. Patriarche, L. Liu, J.-C. Harmand, and V. Zwiller, Nano Lett.
  {\bf 10},  1198  (2010).

\bibitem{our-cpqd}
M. {Bouwes Bavinck}, K.~D. J{\"{o}}ns, M. Zieli{\'{n}}ski, G. Patriarche, J.-C.
  Harmand, N. Akopian, and V. Zwiller, Nano Lett. {\bf 16},  1081  (2016).

\bibitem{maaike-strain}
M. Bouwes~Bavinck, M. Zieliński, B.~J. Witek, T. Zehender, E.~P. A.~M.
  Bakkers, and V. Zwiller, Nano Lett. {\bf 12},  6206  (2012).

\bibitem{nw-sharp}
G. Priante, F. Glas, G. Patriarche, K. Pantzas, F. Oehler, and J.-C. Harmand,
  Nano Lett. {\bf 16},  1917  (2016).

\bibitem{zielinski-multiscale}
M. Zieliński, Acta Phys. Pol. A {\bf 122},  312  (2012).

\bibitem{rozanski-zielinski}
P.~T. Różański and M. Zieliński, Phys. Rev. B {\bf 94},  045440  (2016).

\bibitem{rozanski-zielinski-cpc}
P.~T. Różański and M. Zieliński, Comput. Phys. Commun. {\bf 238},  254
  (2019).

\bibitem{tan-fit}
Y.~P. Tan, M. Povolotskyi, T. Kubis, T.~B. Boykin, and G. Klimeck, Phys. Rev. B
  {\bf 92},  085301  (2015).

\bibitem{Bir1974}
G.~L. Bir and G.~E. Pikus, {\em {Symmetry and strain-induced effects in
  semiconductors}} (Wiley, 1974).

\bibitem{Trebin1979}
H.~R. Trebin, U. R{\"{o}}ssler, and R. Ranvaud, Phys. Rev. B {\bf 20},  686
  (1979).

\bibitem{bahder90}
T.~B. Bahder, Phys. Rev. B {\bf 41},  11992  (1990).

\bibitem{Kadantsev2012}
E.~S. Kadantsev and P. {Zieli{\'{n}}ski Micha{\l}and Hawrylak}, Phys. Rev. B
  {\bf 86},  85411  (2012).

\bibitem{Suzuki1974}
K. Suzuki and J.~C. Hensel, Phys. Rev. B {\bf 9},  4184  (1974).

\bibitem{Slater1954}
J.~C. Slater and G.~F. Koster, Phys. Rev. {\bf 94},  1498  (1954).

\bibitem{Jancu1998}
J.-M. Jancu, R. Scholz, F. Beltram, and F. Bassani, Phys. Rev. B {\bf 57},
  6493  (1998).

\bibitem{harrison-big-book}
W.~A. Harrison, {\em Electronic Structure and the Properties of Solids}
  (Freeman, New York, 1980).

\bibitem{mayer91}
H. Mayer and U. R{\"{o}}ssler, Phys. Rev. B {\bf 44},  9048  (1991).

\bibitem{Winkler2003}
R. Winkler, {\em {Spin-Orbit Coupling Effects in Two-Dimensional Electron and
  Hole Systems}} (Springer, 2003).

\bibitem{Luttinger1955}
J.~M. Luttinger and W. Kohn, Phys. Rev. {\bf 97},  869  (1955).

\bibitem{vurgaftman01}
I. Vurgaftman, J.~R. Meyer, and L.~R. Ram-Mohan, J. Appl. Phys. {\bf 89},  5815
   (2001).

\bibitem{bir61}
G.~L. Bir and G.~E. Pikus, Fiz. Tverd. Tela {\bf 3},  3050  (1961).

\bibitem{pryor98b}
C. Pryor, J. Kim, L.~W. Wang, A.~J. Williamson, and A. Zunger, J. Appl. Phys.
  {\bf 83},  2548  (1998).

\bibitem{Kleinman1962}
L. Kleinman, Phys. Rev. {\bf 128},  2614  (1962).

\bibitem{Yu2010}
P. Yu and M. Cardona, {\em {Fundamentals of semiconductors : physics and
  materials properties}} (Springer, 2010).

\bibitem{Niquet2009}
Y.~M. Niquet, D. Rideau, C. Tavernier, H. Jaouen, and X. Blase, Phys. Rev. B
  {\bf 79},  245201  (2009).

\bibitem{Garg2009}
R. Garg, A. H{\"{u}}e, V. Haxha, M.~A. Migliorato, T. Hammerschmidt, and G.~P.
  Srivastava, Appl. Phys. Lett. {\bf 95},  41912  (2009).

\bibitem{Steiger2011}
S. Steiger, M. Salmani-Jelodar, D. Areshkin, A. Paul, T. Kubis, M. Povolotskyi,
  H.-H. Park, and G. Klimeck, Phys. Rev. B {\bf 84},  155204  (2011).

\bibitem{Mielnik-Pyszczorski2018b}
A. Mielnik-Pyszczorski, K. Gawarecki, and P.~G. M. M. P. G. M. M.~P.
  {Gawe{\l}czyk Micha{\l}and Machnikowski}, Phys. Rev. B {\bf 97},  245313
  (2018).

\bibitem{dyakonov86a}
M.~I. D'yakonov, V.~A. Marushchak, V.~I. Perel, and A.~N. Titkov, Zh. Eksp.
  Teor. Fiz. {\bf 90},  1123  (1986).

\bibitem{Jancu2007}
J.-M. Jancu and P. Voisin, Phys. Rev. B {\bf 76},  115202  (2007).

\bibitem{birner11}
S. Birner, Ph.D. thesis, Technical University of Munich, 2011.

\bibitem{petsc}
S. Balay, J. Brown, K. Buschelman, W.~D. Gropp, D. Kaushik, M.~G. Knepley,
  L.~C. McInnes, B.~F. Smith, and H. Zhang, {{\{}PETSc{\}} {\{}W{\}}eb page},
  2013.

\bibitem{Lee2004}
S. Lee, F. Oyafuso, P. von Allmen, and G. Klimeck, Phys. Rev. B {\bf 69},
  45316  (2004).

\bibitem{zielinski-tallqd}
M. Zieliński, Phys. Rev. B {\bf 88},  115424  (2013).

\bibitem{Note2}
Note reversed ordering of hole states with respect to the electron.

\end{thebibliography}
	
\end{document}